# Multiple Time Series Fusion Based on LSTM: An Application to CAP A Phase Classification Using EEG

Fábio Mendonça[a,b*], Sheikh Shanawaz Mostafa[a], Diogo Freitas[a,c,d], Fernando Morgado-Dias[a,d], and Antonio G. Ravelo-García[e,a]

[a] Interactive Technologies Institute (ITI/ARDITI/LARSyS and M-ITI), 9020-105 Funchal, Portugal.

[b] University of Madeira, 9000-082 Funchal, Portugal.

[c] NOVA Laboratory for Computer Science and Informatics, 2829-516 Caparica, Portugal.

[d] Faculty of Exact Sciences and Engineering, University of Madeira, 9000-082 Funchal, Portugal.

[e] Institute for Technological Development and Innovation in Communications, Universidad de Las Palmas de Gran Canaria, 35001 Las Palmas de Gran Canaria, Spain.

[*] Corresponding author.

E-mail addresses: fabio.mendonca@tecnico.ulisboa.pt (F. Mendonça), sheikh.mostafa@tecnico.ulisboa.pt (S.S. Mostafa), diogo.freitas@m-iti.org (D. Freitas), morgado@uma.pt (F. Morgado-Dias), antonio.ravelo@ulpgc.es (A.G. Ravelo-García).

**Abstract**

Biomedical decision making involves multiple signal processing, either from different sensors or from different channels. In both cases, information fusion plays a significant role. A deep learning based electroencephalogram channels' feature level fusion is carried out in this work for the electroencephalogram cyclic alternating pattern A phase classification. Channel selection, fusion, and classification procedures were optimized by two optimization algorithms, namely, Genetic Algorithm and Particle Swarm Optimization. The developed methodologies were evaluated by fusing the information from multiple electroencephalogram channels for patients with nocturnal frontal lobe epilepsy and patients without any neurological disorder, which was significantly more challenging when compared to other state of the art works. Results showed that both optimization algorithms selected a comparable structure with similar feature level fusion, consisting of three electroencephalogram channels, which is in line with the CAP protocol to ensure multiple channels' arousals for CAP detection. Moreover, the two optimized models reached an area under the receiver operating characteristic curve of 0.82, with average accuracy ranging from 77% to 79%, a result which is in the upper range of the specialist agreement. The proposed approach is still in the upper range of the best state of the art works despite a difficult dataset, and has the advantage of providing a fully automatic analysis without requiring any manual procedure. Ultimately, the models revealed to be noise resistant and resilient to multiple channel loss.

**1. Introduction**

Information fusion technologies enable the combination of information from multiple sources in order to unify and process data. These technologies can thus transform the information from different sources into a representation that provides effective support for automatic analysis [1]. In essence, there are two fundamental methods to process the information from multiple sources. The first, known as centralized fusion, employs a fusion center to receive and process all the information from the different sources, while in the second, known as distributed fusion, each source provides a local estimation from its measured data to the fusion node which then performs the fusion. The first method can attain optimal performance. However, the second has a higher robustness, a relevant characteristic especially when biomedical sensors, such as electroencephalogram (EEG), are used since these can be easily contaminated with noise or can lose contact [2].

Information fusion was applied with success in numerous fields [3], among these, body sensors' analysis attained significant developments with revolutionary applications in health-care and fitness examination [4]. The fusion of information from multiple sources allows the reduction of noise effects, improves the robustness against interference, and reduces ambiguity and uncertainty, seeing that the use of an individual source of information is often not sufficient to provide a reliable examination.

The hierarchy of information fusion can be divided in three main levels. First is the data level fusion techniques, such as Kalman filter and averaging methods, operating at the lowest level of abstraction to combine raw data from multiple sources [5]. The second performs the fusion at the feature level, where feature sets extracted from different data sources are combined to create a new feature vector. The last one is carried out at the decision level and deals with the selection (or creation) of an hypothesis from the set of hypotheses, and is usually performed by fuzzy logic, Bayesian inference, classical inference, or heuristic-based schemes (such as majority voting) [4]. The data level and feature level fusion are generally done before classification or any hypothesis selection or creation about the data. Afterwards, the decision level fusion is done.

Cyclic Alternating Pattern (CAP) is characterized by sequences of transient electrocortical events in the brain, divergent from the background activity. The CAP concept can be used to examine the sleep microstructure during the non-rapid eye movement sleep and is composed of an initial phase of brain activation, named A phase, followed by a period of return to the background activity, denoted B phase. Both phases must have a duration between two and 60 seconds to be considered valid. Two or more successive CAP cycles define a CAP sequence [6] [7] [8].

The activity in the brain is identified using different signals such as the ones coming from the EEG channels [5]. In this view, A phase classification is a suitable problem for fusion based approaches. Therefore, it was hypothesized in this work that the fusion of multiple EEG channels could provide more relevant information for the automatic A phase classification when compared to single-channel models. In other words, the main goal of this work is to develop an automatic classifier for the A phase assessment based on the signals from multiple EEG channels.

CAP has shown to be related to formation, consolidation, and disruption of the sleep macrostructure, working as a measure of the brain's effort to maintain sleep [8] [9] [10]. It was also acknowledged as an EEG marker of sleep instability and a temporal relationship between the CAP, behavioral activities, and autonomic functions was observed [10]. Hence, the CAP was found to be linked with the incidence of several sleep disorders including insomnia [11], Nocturnal Frontal Lobe Epilepsy (NFLE) [12], sleep apnea [13], periodic limb movements [14], and idiopathic generalized epilepsy [15].

Therefore, the employment of CAP analysis by the sleep centers can lead to significant advances in the diagnosis and characterization of sleep quality. However, the introduction of CAP analysis as a regular clinical practice faces some obstacles, namely (a) the time required for manually scoring a whole night polysomnography (the gold standard for sleep analysis [16]), due to the large amount of information produced during whole night EEG recording, (b) combining different information from different sensors or channels, (c) the need for qualified personnel to perform the manual scoring, and (d) the fair inter-scorer specialist agreement, that varies from 69% to 78% [17].

Therefore, manual scoring is considerably problematic, as the process is unpractical and prone to misclassifications. For these reasons, the development of algorithms for automatic CAP analysis with information fusion is desirable, supporting, thus, the necessity for this study.

It was also observed that CAP is a global EEG phenomenon that comprises extensive cortical areas, suggesting that the A phases could be visible on all EEG channels [6]. However, the state of the art works which proposed methodologies for automatic A phase analysis perform the examination using only one EEG channel (usually with one monopolar derivation). Although this approach can lead to less complex models, it is also reductive and restrictive since a large amount of information coming from the other channels is discarded, disregarding at the same time the fact that the A phase activity can occur over multiple cortical areas.

In addition, most of the methods proposed in the state of the art for A phase detection employ classification with features created by the researchers. Nevertheless, significant domain-specific knowledge is required for the feature creation process and it is becoming increasingly challenging to discern a new set of features that can reach a better performance than the methods already reported in the state of the art. Also there is the need for feature sorting which does not guarantee a performance improvement [18] [19]. These complications can be surpassed by a deep learning model, which can automatically learn the relevant patterns from the input signal. However, a significant gap in the state of the art regarding deep learning applications for CAP analysis was identified.

CAP phases have a strong temporal dependency that can be captured by recurrent neural networks, e.g. Long Short-Term Memory (LSTM) [20], and the activity can be measured in different EEG channels. Therefore, a novel approach was followed in this work where the information from multiple EEG channels was fused by a proposed deep learning channel fusion methodology, composed of LSTM, concatenation, and fully connected (dense) layers.

On the other hand, the structure and/or hyperparameters of a deep learning classifier are usually selected through an experimental search (usually a grid search), which performs an exhaustive evaluation of multiple combinations of parameters. However, this approach requires a significant amount of time and computational resources, which can be impracticable for deep learning models [21]. Two heuristic based algorithms, namely, Genetic Algorithms (GA) and Particle Swarm Optimization (PSO), were used in this work as an alternative to the grid search approach to find the optimal structure, number of channels, and hyperparameters of the models [18]. Therefore, two models were developed to perform the channel fusion of EEG channels for the CAP A phase assessment, one was tuned by a GA and the other by the PSO. It is also intended to study the optimization algorithms characteristics to determine which can lead to the best performance.

The key novelties of this work can be summarized as follows:

-Proposal of a novel method for information fusion based on a deep learning model which is responsible for extracting the features, performing the feature level fusion, and performing the classification. The structure of the classifier was tuned by the optimization algorithm hence, all the fusion and classification procedures were optimized and performed automatically by the deep learning model which learned the relevant patterns directly from the data.

-Independent evaluation of two optimization algorithms for finding the optimal structure of a deep learning classifier. The optimization of deep learning models is a well-known difficulty in the machine learning field since the simulations are usually slow, hence, there is a need to study suitable algorithms to haste this process.

-Combined examination of subjects free from neurological disorders and subjects with a sleep related disorder using information (the signal) from multiple EEG channels to assess the CAP A phases. The state of the art standard is to only examine one channel for the analysis, which is contrary to the specification of the CAP protocol where it specifies that the analysis should preferably be carried out over multiple channels.

- Development of systems tolerant to noise (until a signal to noise ratio of 0 dB) and able to handle the loss of 66% of the information, i.e. loss of two channels.

It is important to highlight here that the CAP A phase assessment was used as an example of the application of the proposed fusion of multiple time series. This means that the suggested approach was developed to be generic, and thus be applied to other research and industry applications.

The paper has the following organization: an overview of the state of the art for CAP A phase analysis is presented in Section 2; the employed materials and methods are presented in Section 3; the model's performance is evaluated in Section 4; a discussion of the obtained results is carried out in Section 5; the paper is concluded in Section 6.

**2. Overview of the state of the art for CAP A phase analysis**

A literature review was conducted based on the PRISMA style, covering papers published until June 2021. The search was conducted using the IEEE Xplore, PubMed, Web of Science, and cited literature in the included articles. The keywords used in the search were "cyclic alternating pattern" and "CAP AND A phase AND EEG". The inclusion criterion was the presentation of a method for the A phase detection while the exclusion criterion was the absence of a performance metric that can advocate the capability of the model. The search was performed on articles written in English. The search yielded 2420 and 239 results for the first and second keywords string, respectively. After removing the duplicated and the non-relevant articles, the total of selected articles were 26, published from 2002 to 2021.

Two main research lines were found in the state of the art regarding the CAP analysis. The first comprises the evaluation of the A phase subtypes [22] [23] [24] [25] [26] [27] [28] [29] [30] [31] while the second performs the A phase detection for the CAP cycle estimation, usually considering each epoch as either "A" or "not-A", leading to a binary classification problem. Although the A phase subtypes can provide significant information about the sleep process, this information is not required for the sleep stability assessment based on CAP cycle analysis since it only considers the occurrence of an activation and not the subtype. Therefore, the research line followed by this work comprises the A phase binary classification. A total of 16 articles were found performing this analysis.

Largo et al. [32] [33] examined the power of five EEG frequency bands (delta, theta, alpha, sigma, and beta). The fast discrete wavelet transform was applied to the signal of one EEG channel, and evaluated two moving averages with a short and a long duration. The relationship between the averages was named as activity index and it was used to detect the occurrence of A phases. The classification model was tuned by a GA, and the A phases assessment was performed by comparing with a threshold. Niknazar et al. [34] employed a classification method based on a similarity analysis between the windowed input signal and reference windows from a database, evaluating the signal from one EEG monopolar derivation (C4–A1 or C3–A2). Barcaro et al. [35] employed a technique to describe the sleep microstructure by computing band descriptors, one for each of the five EEG characteristic bands evaluated in the F4–C4 channel, to measure how much the amplitude of a frequency band differs from the background activity. Afterwards, a tuned threshold was used to implement the classification.

Mariani et al. [36] also employed a threshold based classification, evaluating the differential variance of the EEG signal from one monopolar derivation (C4–A1 or C3–A2), Hjorth descriptors (activity and mobility, computed in the low delta and high delta bands), and the band descriptors. The highest accuracy was attained by the first feature. These features were also used by Mariani et al. [37], [38], [39], and [40]. A Forward Neural Network (FFNN) was used in the first work [37] examining the signal from one monopolar derivation (C4–A1 or C3–A2), while a Support Vector Machine with Gaussian kernel was employed in the second [38], evaluating the F4–C4 channel. Four classifiers were tested in the third work [39] and the Linear Discriminant Analysis (LDA) attained the highest accuracy, when evaluating the EEG signal from one monopolar derivation (C4–A1 or C3–A2). A variable window was employed in the fourth work [40] for the feature creation process using an EEG monopolar derivation (C4–A1 or C3–A2), then the features were fed to a different discriminant function for each A phase subtype. Afterwards, the outputs were combined for the A phase classification.

Mendonça et al. [41] and [42] evaluated the Shannon entropy, log-energy entropy, Teager Energy Operator (TEO), auto-covariance, standard deviation, and power spectral density of the five characteristic EEG frequency bands. LDA was employed in the first work [41] to perform the classification while multiple classifiers were examined in the second work [42]. It was concluded that FFNN attained the best accuracy. In both works, the signal from one EEG monopolar derivation (C4–A1 or C3–A2) was examined.

Sharma et al. [43] applied a five level wavelet decomposition to produce six sub-bands and four features were extracted for each sub-band, specifically, the three Hjorth parameters (activity, mobility, and complexity) and the wavelet entropy. These features were then fed to an ensemble of bagged tree classifier. The examination was performed by extracting features from two channels, one monopolar (C4–A1) and one bipolar (F4–C4) EEG derivations.

Mostafa et al. [44] proposed a deep learning method to classify each two second window with a deeply-stacked auto encoder fed with the signal from one EEG monopolar derivation (C4–A1 or C3–A2). Hartmann and Baumert [45] examined the same EEG derivation, feeding entropy and frequency based features, differential variance, and TEO to an LSTM to the classification procedure. Mendonça et al. [20] and [46] also employed an LSTM. However, the classifier was directly fed with the EEG signal from one monopolar derivation (C4–A1 or C3–A2).

From the state of the art analysis, it was concluded that tunable thresholds were frequently used at the beginning of the automatic CAP analysis which then were later replaced by the features based machine learning classification. The most recent articles employ deep learning methods, pointing to the occurrence of a new trend where the feature creation process is performed automatically by the classifier. Another relevant aspect is that only the monopolar derivations and the F4–C4 channel were previously studied by state of the art works. The summary of the state of the art analysis is presented in Fig 1.

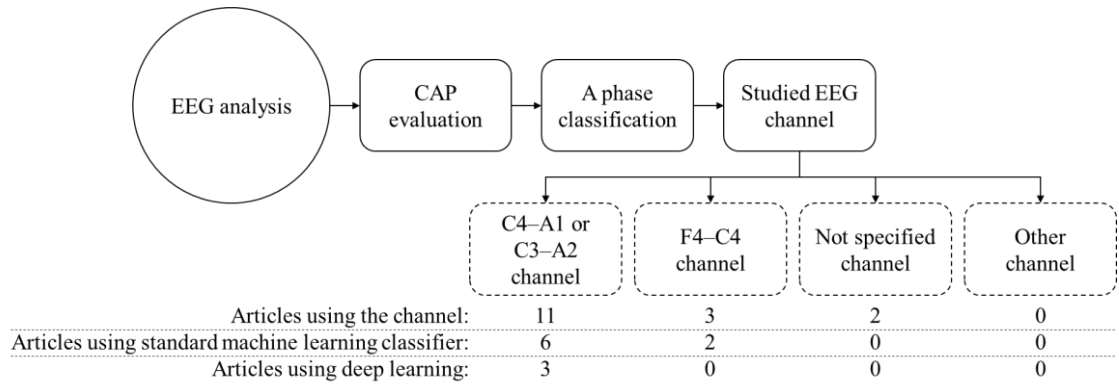

**Fig. 1.** Overview of the state of the art examination.

## 3. Materials and methods

The developed model estimates the CAP A phases, in a second-by-second assessment, by examining the pre-processed signals from multiple EEG channels.

Those signals were fused by the deep learning classifier which performed the automatic feature extraction and classification. Specifically, distributed fusion was employed in this work since it is suitable when the sources of information come from similar sensors [47]. Each EEG channel was fed to one LSTM which was used to extract features from each signal. Afterwards, the extracted features were concatenated by the fusion node to produce the fused feature vector (feature level fusion [4]) employed to perform the A phase classification.

The classifier's output was then post-processed to diminish the misclassification, and the model's performance was assessed. Two optimization algorithms were examined to tune the model with the goal of finding the optimal number of channels, number of time steps for the classification's recurrence, and structure of the classifier. The pseudocode of the developed model is presented in Algorithm 1. The code developed for this work was made open-source, being publicly available in a GitHub repository (https://github.com/Dntfreitas/GA_PSO_DEEP_LEARNING).

**Algorithm 1.** Pseudocode for the experimental procedure.

1: *Input*: $EEG_{channels}$ ← {Fp2–F4, C4–A1, F4–C4}
2: **For each** *ch* **in** *EEG_{channels}* **do**:
3:     *ch* ← *resample*(*ch*)      ⎱ Preprocessing ► Sec. 3.5
4:     *ch* ← *standardize*(*ch*)      ⎰
5: *optimizers* ← [GA, PSO]      ► Sec. 3.3.1, Sec. 3.3.2
6: **For each** *opt* **in** *optimizers* **do**:
7:     **While** *opt*'s termination criteria **is not** met **do**:
9:        Initialize the deep learning models' structure
10:       Fit each deep learning model:      ► Sec. 3.5
11:         Feature extraction using LSTM with or without a dense layer    ⎱ Feature extraction and fusion ► Sec. 3.2
12:         Fusion node implemented by the concatenation layer    ⎰
13:         Classify A phases using the classifying dense layer    ⎱ Classification ► Sec. 3.2
14:         $O \leftarrow [O_0, O_1, ..., O_{ep-1}]$, where $O_{ep}$ is the output of the $ep^{\text{th}}$ epoch of the deep learning model
            $T \leftarrow O$
15:         **For each** output *out* = 0, ..., *ep* − 1 **do**:    ⎱ Post-processing ► Sec. 3.5
16:           **If** *out* > *0* **and** *out* < *length*(*O*) **do**:
               $T_{out} = majority(O_{out-1}, O_{out}, O_{out+1})$
        $O \leftarrow T$
18:        Performance assessment      ► Sec. 3.4
19: Find the best structure for GA and PSO for the deep learning model, using *PM* as reference      ► Sec. 4.1
20: Performance comparison between GA and PSO      ► Sec. 4.2
21: Robustness evaluation for noise and channel failure      ► Sec. 4.3

▲ Optimization procedure Sec. 3.3

### 3.1 Studied population

Recordings from the CAP Sleep Database [6] [48] were selected to develop the model. This database is publicly available and has annotations provided by sleep experts regarding the A phase occurrence and duration. Relevant information for the CAP analysis is present in both EEG bipolar and monopolar derivations since the CAP is a global EEG phenomenon, comprising broad cortical areas [6]. As reported by Mariani et al. [38], CAP analysis is usually performed using only the signal from one monopolar derivation (either C4–A1 or C3–A2). However, such methodology is prone to have a large number of false positives (identified A phases) as many activations correspond to changes in amplitude and/or frequency on the central lead but are regular EEG rhythms on the others. Therefore, CAP scoring should be performed by scoring multiple channels [38].

In that view, the goal is to use as many derivations as possible while keeping the model's complexity feasible to be used in the current available hardware. It was observed that the state of the art works examined either the F4–C4 channel or one monopolar derivation (C4–A1 or C3–A2). Nevertheless, Terzano et al. [6] indicated that all bipolar derivations can properly detect the A phases, hence, the Fp2–F4 was also examined in this work. Hence, the three examined deviations are Fp2–F4, F4–C4, and C4–A1.

Eight subjects Free of Neurological Disorder (FND) were chosen from the dataset since these were the ones that have the examined channels available. To provide a broader representation of the general population, eight more subjects having the same three EEG deviations with a sleep related disorder (also available in the dataset) were included. NFLE was chosen to be the studied disorder since the epileptic manifestations are likely to act as a subcontinuous "internal noise" which can induce a substantial growth of all CAP related parameters, reflecting the degree of sleep instability [12]. According to our best knowledge, no state of the art work examined a combination of normal subjects and subjects with NFLE in the task of automatic classification of CAP A phases.

Therefore, the considered population was composed of eight normal subjects (reference for normal sleep quality) and eight subjects prone to have poor sleep quality.

A population of 16 subjects (eleven females and five males) was found to be either equal or higher than the works available in the state of the art performing the CAP A phase analysis.

The average total sleep time of the studied population was 463.97 minutes, ranging from 370.5 to 553.5 minutes, with a standard deviation of 54.21 minutes. The average subject's age was 32.88 years old, ranging from 16 to 67 years old, with a standard deviation of 11.43 years old. The number of one second epochs related to the occurrence of an A phase was 67118 and the total number of one second epochs was 518723.

**3.2 Classification and channel fusion**

The information fusion concept was employed to combine data from multiple EEG channels [4]. This fusion was performed at the feature level where the multiple feature vectors were combined to form the joint feature vector from which the classification was performed. The features were automatically created for each EEG channel by feeding the pre-processed signal to the designated LSTM layer for the channel.

Each LSTM layer is composed of memory cells which sequentially process the input and preserve their hidden state through time [49]. Each cell is controlled by three gates. The input gate ($I$) defines the flow of activations into the cell while the output gate ($O$) controls the flow of activations to the remaining network. The forget gate ($F$) is responsible for adaptively resetting the cell's state. For the time step $t$ and cell $c$, these operations are defined as [50]

$$F_c^{(t)} = \sigma \left( \sum_j U_{c,j}^F x_j^{(t)} + \sum_j W_{c,j}^F h_j^{(t-1)} + b_c^F \right) \quad (1)$$

$$I_c^{(t)} = \sigma\left(\sum_j U_{c,j}^I x_j^{(t)} + \sum_j W_{c,j}^I h_j^{(t-1)} + b_c^I\right) \quad (2)$$

$$O_c^{(t)} = \sigma\left(\sum_j U_{c,j}^O x_j^{(t)} + \sum_j W_{c,j}^O h_j^{(t-1)} + b_c^O\right) \quad (3)$$

where $\sigma$ is the sigmoid function given by $\sigma(\alpha) = 1/(1+e^{-\alpha})$, $\boldsymbol{x}^{(t)}$ is the input vector, $\boldsymbol{U}$ are the input weights, $\boldsymbol{W}$ are the recurrence weights, and $\boldsymbol{b}$ are the bias. The network's output, $\boldsymbol{h}$, is given by [50]

$$h_c^{(t)} = tanh\left(s_c^{(t)}\right)o_c^{(t)}$$

where $tanh$ is the hyperbolic tangent function calculated as $tanh(\alpha)=2\sigma(2\alpha)–1$, and $s^{(t)}$ is the cell's internal state, updated by

$$s_c^{(t)} = f_c^{(t)} s_c^{(t-1)} + i_c^{(t)} tanh\left(\sum_j U_{c,j} x_j^{(t)} + \sum_j W_{c,j} h_j^{(t-1)} + b_c\right) \quad (4)$$

An LSTM layer can examine the data sequence in only one direction (conventional LSTM model) or in two directions, denoted as Bidirectional LSTM (BLSTM). Although the BLSTM models use more parameters when compared to the conventional LSTM models, it is likely that these models can find more relevant patterns on the fed data.

Each LSTM cell receives a time step of data with duration $D$, composed of $I$ input points. The type of LSTM, the number of channels, $n$, number of time steps, $T$, and number of LSTM layers (stacked if more than one) were chosen by the optimization algorithm. Each cell has multiple hidden units and the total number of hidden units, $H$, of the last cell defines the output of the LSTM layer (the epoch's data fed to the last cell corresponds to the database label for the current evaluated epoch). When two LSTM layers were stacked, the sequence of vectors of the first layer was returned to the second layer whose last cells' outputs defined the output.

The LSTM layers' outputs the features $h_1, h_2, \ldots, h_n$ that were automatically crafted from each input channel. These features were then transformed to $f = [f_1 [h_1(1), h_1(2), \ldots, h_1(H)], f_2 [h_2(1), h_2(2), \ldots, h_2(H)], \ldots, f_n [h_n(1), h_n(2), \ldots, h_n(H)]]$ by the concatenation layer, where $f_1, f_2, \ldots, f_n$ are either the outputs of the LSTM ($h_1, h_2, \ldots, h_n$), or are the dense layers' transformations of the LSTM outputs, according to the decision of the optimization algorithm. These channels were fused, at the feature level, by the concatenation layer which merges all the features into a sequence $f$, i.e. the input of the fusion node is the set of features $h$ and the output is $f$. If a dense layer was used to transform the LSTM layer's outputs then, a second dense layer (with the same configuration as the first dense layer) was used to transform the concatenation layer's output.

At the end, the softmax function, given by $\text{softmax}(\alpha) = e^{\alpha} / \sum_j e^{\alpha_j}$, was used by a fully connected layer to normalized the output. At the end, binary classification output was obtained by applying the max operation.

### 3.3 Optimization procedure

Two optimization algorithms were studied to find the best structure of the classifier for the A phase assessment, evaluating an encoding array. These stochastic algorithms were used in this work as an alternative to the conventional grid search, which is considered unfeasible especially when many parameters have to be tuned [18]. The pseudocodes for GA and PSO are presented in algorithms 2 and 3, respectively, where the goal is to find the solution which maximizes the Performance Metric (*PM*).

The GA was selected since it is one of the most commonly used algorithms for complex design optimization problems, using Darwinian principles of biological evolution [18]. On the other hand, PSO methodology is based on information sharing, such as occurs in nature in the flocks of birds and schools of fish, and this algorithm was selected since it has large flexibility and is capable of finding the globally best solution in complex (possibly multimodal) search spaces [51].

**Algorithm 2.** Pseudocode for the GA variant used in this work.

1:   $g \leftarrow 0$, $Pa \leftarrow 0$
2:   $m_{prob}^{(g)} \leftarrow 0.2$, $c_{prob} \leftarrow 0.9$
3:   $best_{fit} \leftarrow -\infty$
4:   Randomly initialize $P^{(g)}$ with $z$ chromosomes
5:   Evaluate the chromosomes in $P^{(g)}$
6:   $P^{(g)} \leftarrow$ *sort descending* $(P^{(g)})$
7:   **While** $g < G$ and $Pa < Pa_{max}$ **do**:
8:     $g \leftarrow g + 1$
9:     $m_{prob}^{(g)} \leftarrow m_{prob}^{(0)} - m_{prob}^{(0)} \times \lfloor g/5 \rfloor \times 0.3$
10:    **If** $m_{prob}^{(g)} < 0.01$ **do**:
11:       $m_{prob}^{(g)} \leftarrow 0.01$
12:    Initialize $Q$ with $z$ chromosomes
13:    **For each** chromosome $q$ in $Q$ **do**:
14:      **If** $rand() \leq c_{prob}$ **do**:
15:        **While** $parent_1 = parent_2$ **do:**
16:          $parent_1 \leftarrow min(\{p_{rand()} \in P^{(g-1)}, p_{rand()} \in P^{(g-1)}\})$
17:          $parent_2 \leftarrow min(\{p_{rand()} \in P^{(g-1)}, p_{rand()} \in P^{(g-1)}\})$
18:        $q \leftarrow crossover(parent_1, parent_2)$
19:      **Else:**
20:        $q \leftarrow min(\{p_{rand()} \in P^{(g-1)}, p_{rand()} \in P^{(g-1)}\})$
21:      **For each** bit $b$ in $q$ **do**:
22:        **If** $rand() \leq m_{prob}^{(g)}$ **do:**
23:          $b \leftarrow \sim b$
24:    $P_0^{(g)} \leftarrow P_0^{(g-1)}$
25:    $P_1^{(g)} \leftarrow P_1^{(g-1)}$
26:    Evaluate the chromosomes in $Q$

(braces on right: Crossover, Cloning, Mutation, Elitism)

27:     $T \leftarrow \textit{sort descending}([P^{(g-1)}, Q])$
28:     $P_{2:z}^{(g)} \leftarrow T_{0:z-2}$
29:     $P^{(g)} \leftarrow \textit{sort descending}(P^{(g)})$
30:     **If** $PM(P_0^{(g)}) > best_{fit}$ **do:**
31:        $best_{fit} \leftarrow PM(P_0^{(g)})$
32:        $Pa \leftarrow 0$
33:     **Else:**
34:        $Pa \leftarrow Pa + 1$

**Algorithm 3.** Pseudocode for the PSO algorithm variant used in this work.

1:   $i \leftarrow 0$, $Pa \leftarrow 0$
2:   $\omega^{(i)} \leftarrow 0.9$, $c_1 \leftarrow 0.6$, $c_2 \leftarrow 0.3$
3:   $best_{fit} \leftarrow -\infty$
4:   Randomly initialize $x^{(i)}$ and $v^{(i)}$, for every particle in $S^{(i)}$
5:   Evaluate the particles in $S^{(i)}$
6:   $p^{(i)} \leftarrow x^{(i)}$, for every particle in $S^{(i)}$
7:   Update $l^{(i)}$, for every particle in $S^{(i)}$ according to the neighboring particles
8:   **While** $i < G$ **and** $Pa < Pa_{max}$ **do**:
9:     $i \leftarrow i + 1$
10:     $\omega^{(i)} \leftarrow \omega^{(0)} - \omega^{(0)} \times \lfloor i/5 \rfloor \times 0.09$
11:     **If** $\omega^{(i)} < 0.4$ **do**:
12:       $\omega^{(i)} \leftarrow 0.4$
13:     **For each** particle $k$ in $S^{(i)}$ **do**:
14:       **For each** bit $b$ in $k$ **do**:
15:         $r_1 \leftarrow rand()$
16:         $r_2 \leftarrow rand()$
17:         $v_{k,b}^{(i)} \leftarrow \omega^{(i-1)} \times v_{k,b}^{(i-1)} + c_1 \times r_1(p_{k,b}^{(i-1)} - x_{k,b}^{(i-1)}) + c_2 \times r_2(l_{k,b}^{(i-1)} - x_{k,b}^{(i-1)})$
18:         **If** $rand() < \sigma(v_{k,b}^{(i)})$ **do**:
19:           $x_{k,b}^{(i)} \leftarrow 1$
20:         **Else**:
21:           $x_{k,b}^{(i)} \leftarrow 0$
22:     Evaluate the particles in $S^{(i)}$
23:     **For each** particle $k$ in $S^{(i)}$ **do**:
24:       **If** $PM(x_k^{(i)}) > PM(p_k^{(i-1)})$ **do**:
25:         $p_k^{(i)} \leftarrow x_k^{(i)}$
26:       **Else**:
27:         $p_k^{(i)} \leftarrow p_k^{(i-1)}$
28:     Update $l^{(i)}$, for every particle in $S^{(i)}$ according to the neighboring particles
29:     **If** $max(\{PM(p_k^{(i)}) \mid k \in S^{(i)}\}) > best_{fit}$ **do**:
30:       $best_{fit} \leftarrow max(\{PM(p_k^{(i)}) \mid k \in S^{(i)}\}$
31:       $Pa \leftarrow 0$
32:     **Else**:
33:       $Pa \leftarrow Pa + 1$

Lines 15–17: Velocity. Lines 18–21: Position. Lines 22–28: Cognitive and social information.

### 3.3.1 Genetic algorithm

GA is a type of metaheuristic algorithm that has previously shown to be capable of finding an improved solution over time by replicating the best solutions from generation to generation and producing offspring from these solutions [52].

For this work, the algorithm was initialized with a random individual generation, using mutation and crossover operators over a defined number of generations to reach a solution, which optimized the model to a given metric.

Coded chromosomes were employed to characterize the population $P = [p_1, p_2, …, p_z]$, where $z$ is the size of each generation, $g$. Each $p$ was decoded using a decoding table (see section 3.5), and the quality of the solution (fitness assessment) was assessed by the selected *PM*.

The algorithm stopped if the maximum number of generations, $G$, or if the patience value, $Pa$, (number of consecutive generations that the algorithm did not produce an improved solution) reached the maximum patience, $Pa_{max}$. The initial population of $P$ was randomly generated and then sorted according to the performance of each chromosome. Afterwards, a new cycle started for the creation of the offspring population, $Q$, with size $z$. According to the crossover probability, each new member of the offspring population, $q$, was created either by a two-point crossover operation between two different elements randomly chosen from $P$, or by cloning the most fitted element selected from a tournament of two. In the two-point crossover operation, each crossover produced one offspring, and each of the elements of $P$ can be chosen to participate in a tournament of two, implementing the no-replacement tournament selection [53]. The approach chooses the most fitted element of each tournament to produce the cross-over without allowing the same chromosome to be the winner of the two tournaments, since the tournaments are repeated until two different elements of $P$ are selected. This two-point crossover approach was adopted because it was reported that it outperforms other conventional crossover operations [54]. It is important to note here that in both cases, all the chromosomes have an equal probability of being picked for a tournament, i.e.,

$$\frac{2z - 1}{z(z - 1)}, \text{if and only if}, z \geq 3 \tag{5}$$

The most fitted elements will, however, have a higher probability of being selected in each tournament, and consequently, used for crossover or cloning.

A mutation operation (that performs the logical not operation) was applied to all elements of the chromosome of each $q$ according to the mutation probability, $m_{prob}$. Therefore, the estimated number of mutations on a given iteration $g$ is given by

$$m_{prob}^{(g)}(z - 2)N_{bits} \tag{6}$$

where $N_{bits}$ is the number of bits used to encode the problem. The implemented methodology for the GA follows the convention of starting with high exploration (using a high $m_{prob}$) and then progressively changing into exploitation (decreasing $m_{prob}$ every five generations). It is worth noting that if both mutation and crossover rates are too high, then the GA will head toward random search, while the opposite leads to a hill climbing algorithm. Hence the gradual change from exploration to exploitation is more suitable [55].

The two best $p$ of each generation were considered elites ensuring that they were moved to the next generation. Subsequently, the performance of each $q$ was assessed and stored. $P$ (without the two elites) and $Q$ were combined and sorted according to the performance scores (attained $PM$ by the model defined by the chromosome), from most to least fitted, and the best $z - 2$ members were chosen to compose the new $P$. Afterwards, the two elites were introduced in $P$ which was then sorted from most to least fitted (according to the performance scores). Subsequently, a new generation started and the process was repeated until either *g* was equal to *G* or *Pa* was equal to *Pa$_{max}$*.

### 3.3.2 Particle swarm optimization

PSO is a population-based stochastic optimization algorithm that uses agents (called particles), organized in a swarm (*S*), in order to search for the optimal solution(s) in a (possibly complex) search space. Each particle *p*, in its turn, is a candidate solution for the optimization problem at hand.

The algorithm was initially proposed in 1995 by R. Eberhart and J. Kennedy [56][57]. These authors suggested a collective search strategy in which particles consider the best position found by the other particles (in other words, the social information) and its individual best position (also known as the cognitive information) in order to explore the search space and converge to the optimal solution(s).

In short, PSO can be described in three main steps: (i) initialize the swarm by randomly positioning the particles in the search space; until a stopping criterion is met: (ii) compute, for each particle, its new velocity (*v*) and position (*x*), and (iii) for each particle, when a better solution is found, update the cognitive and social position information.

It is important to note here that the social position information is shared using information links between particles. These information links allow particles to be fully connected, and thus share information with every particle in the swarm or create neighbors of particles where the knowledge is restricted to the particles that belong to the same neighborhood.

In order to optimize the structure and hyperparameters of the deep learning classifier used in this work, a discrete binary PSO [58] variant was used. The velocity of a particle, at every iteration *i* and dimension *d*, was thus updated as follows

$$v_d^{(i+1)} = \omega^{(i)} v_d^{(i)} + c_1 r_1^{(i)} \left( p_d^{(i)} - x_d^{(i)} \right) + c_2 r_2^{(i)} \left( l_d^{(i)} - x_d^{(i)} \right) \qquad (7)$$

where $\omega$ is the inertia weight parameter [59], $c_1$ and $c_2$ the cognitive and social weight respectively, and $r_1$ and $r_2$ two uniformly distributed pseudorandom numbers. Finally, $p$ is the personal best position found by the particle and $l$ the best position found by the neighboring particles. After computing the velocity of the particles, the position of each particle is changed according to

$$x_d^{(i+1)} = \begin{cases} 1, & \text{if } rand() < \sigma\left(v_d^{(i+1)}\right) \\ 0, & \text{otherwise} \end{cases} \qquad (8)$$

where $rand()$ denotes a pseudorandom number drawn from a uniform distribution on the interval [0, 1] and $\sigma$ the sigmoid function.

The particles were organized in a ring topology, where each particle only shares information with the two immediately adjacent neighborhoods. The rationale behind the choice of this topology has to do with the fact that in a ring topology, the social information flows slowly, which simultaneously slows down the convergence speed. This behavior is particularly important in multi-modal complex optimization problems like the one presented in this paper. Having a low convergence rate improves the algorithm's exploration capabilities, prevents the premature convergence of the algorithm and, therefore, reduces the susceptibility of PSO to getting trapped in a local minimum [60] [61]. The inertia weight parameter ($\omega$), on the other hand, was updated following a negative non-linear time-varying approach.

### 3.4 Performance metrics and validation methodology

The performance in the experimental results was assessed by the Accuracy (Acc), Sensitivity (Sen), and Specificity (Spe) of the predictions against the ground truth (database labels) by [62]

$$Acc = \frac{TP + TN}{TP + TN + FP + FN} \tag{9}$$

$$Sen = \frac{TP}{TP + FP} \tag{10}$$

$$Spe = \frac{TN}{TN + FN} \tag{11}$$

where *TP* is the number of instances of class "A" classified as class "A", *TN* is the number of instances of class "not-A" classified as class "not-A", *FP* is the number of instances of class "not-A" classified as class "A", and *FN* is the number of instances of class "A" classified as class "not-A". The diagnostic ability of the algorithm was evaluated by the Area Under the receiver operating characteristic Curve (AUC) [63], considering that the positive class was "A".

The normalized diversity of the population or particles at each generation or iteration (distance-based measure) was computed as [55] [64]

$$Div(g) = \frac{2}{zL(z-1)} \sum_{\mu=1}^{z-1} \sum_{\theta=\mu+1}^{z} Ham(p_\mu, p_\theta) \tag{12}$$

where *L* is the length of the chromosome or particle, *z* is the number of chromosomes or particles, and *Ham* is the Hamming distance, given by the number of positions where the bits of the two chromosomes differ.

Taking into consideration that the optimization procedure is considerably time consuming, Two-Fold Cross-Validation (TFCV) was used to find the optimized solution with a cold start of the classifier in each run. TFCV was performed by dividing the subjects into two datasets (ensuring subject independent datasets by using the data from each subject exclusively in only one of the datasets). The AUC of the two TFCV cycles was averaged to find the mean AUC considered as the PM for the model under examination. The Adam algorithm [65] was used for training since it was found to be the most suited for the CAP analysis based on LSTM [20]. Cost-sensitive learning was employed to deal with the strong data unbalancement (instead of using a balancing operation that can alter the expected distribution of the data) since for some subjects more than 80% of the epochs can refer to the "not-A" class.

When the best structure of the classifier was found, the Leave One Out (LOO) method was used to assess the performance of that model, with a cold start (the classifier weighs were randomly initialized to not perform retraining) of the classifier in each run. This method was employed as it can provide less biased results when a low number of samples is available [66]. Hence, a total of 16 evaluation cycles were executed. The training set, employed for each cycle, was composed of data from 15 subjects and the data from the left out subject composed the testing set. Each subject was only chosen once to compose the testing set.

**3.5 Implementation**

A resampling procedure was applied to attain a uniform database, since the sampling frequency of the records varies between 100 Hz and 512 Hz. All signals were resampled at the lowest sampling frequency by decimation [67]. A constant reduction factor was employed for the sampling rate, $s$, and a standard lowpass filter (Chebyshev type I filter with order eight, normalized cutoff frequency of $0.8/s$, and passband ripple of 0.05 dB) was used to avoid aliasing and downsample the signal. Thus, a resampling process chooses each $s^{th}$ point from the filtered signal to generate the resampled signal. This signal was then standardized, by subtracting the mean and dividing the result by the standard deviation, with the goal of reducing the effect of systematic variations in the signal [68].

The removal of artifacts related to cardiac field and eye movements during sleep was recommended by several studies as an approach that can marginally improve the performance of the classifier [27] [69]. Nevertheless, the accurate removal of these artifacts requires, at least, the electrooculogram and electrocardiogram signals, leading to a further complex model. Therefore, these artifacts were not removed.

Epoch's duration ($D$) was selected to be one second, which is in line with the standard duration for CAP analysis, and it corresponded to the database labels. Since the signals were resampled at 100 Hz, the input dimension was 100 for each time step.

For this work, AUC was selected as the PM since it can provide an estimate of the diagnostic ability of the algorithm, without being significantly affected by class unbalance. For both studied optimization algorithms, the used learning rate was 0.001 and the batch size was 1024. The optimal classification threshold for the test dataset of the LOO examination was identified by finding the optimal cut off point of the receiver operating characteristic curve estimated on the training dataset.

Four activation functions were assessed by the optimization algorithm to introduce nonlinearities in the network: tanh; sigmoid; Rectified Linear Unit (ReLU); Scaled Exponential Linear Unit (SELU).

An encoding array, presented in Table 1, was employed to perform the optimization search. A total of 15 coded chromosomes or particles, each composed of 15 bits, were employed to characterize the population ($P$ elements) at each generation or iteration, $g$, by using the decoding indicated in Table 1.

**Table 1.**

Encoding array examined by the optimization algorithms.

| Number | Locus | Description | Specification |
|---|---|---|---|
| 1 | 0–2 | Number of channels to be fused | 000: Fp2–F4<br>001: C4–A1<br>010: F4–C4<br>011: Fp2–F4 and C4–A1<br>100: Fp2–F4 and F4–C4<br>101: F4–C4 and C4–A1<br>110 or 111: Fp2–F4, F4–C4, and C4–A1 |
| 2 | 3–4 | Number of time steps to be considered by the LSTM | 00: 10<br>01: 15<br>10: 20<br>11: 25 |
| 3 | 5 | Number of LSTM layers for each channel | 0: One<br>1: Two staked |
| 4 | 6 | Type of LSTM | 0: LSTM<br>1: BLSTM |
| 5 | 7–8 | Shape of the LSTM layers | 00: 100<br>01: 200<br>10: 300<br>11: 400 |
| 6 | 9–10 | Percentage of dropout for the recurrent and dense layers | 00: 0<br>01: 5%<br>10: 10%<br>11: 15% |
| 7 | 11–12 | Size of the dense layers | 00: 0<br>01: 200<br>10: 300<br>11: 400 |
| 8 | 13-14 | Activation function for the dense layers | 00: tanh<br>01: Sigmoid<br>10: ReLU<br>11: SELU |

For GA, the quality of the solution (fitness assessment) for each element of the population was assessed by the average AUC (employed optimization metric since it reveals the diagnostic ability of the model) estimated by TFCV. The values of $G$ and M were chosen to be 20 and 15, respectively. The crossover probability was 90%. The initial mutation probability was 20% and the value was decreased 30% every five generations until the minimum of 1% was reached. The GA parameters were selected to be in line with the ones employed by Largo et al. [32], reported as suitable for CAP analysis using a GA.

To allow a fair comparison with GA, a total of 15 particles were employed with PSO (with the same encoding array defined in Table 1), besides keeping the fitness assessment and the stopping criterion as defined previously for GA. Concerning the specific PSO parameters, $c_1$ was set to 0.6 and $c_2$ to 0.3 in order to lead to the convergence of PSO considering the inertia values [70]. The initial and final values of $\omega$ were defined to be 0.9 and 0.4, respectively [71] [72]. In order to have the same rate of change as the mutation operation in the GA, the value of $\omega$ was decreased by 9% every five generations until the minimum value of 0.4 was reached.

An overview of the implemented model is presented in Fig. 2. Since binary classification was employed, an epoch was considered as misclassified when the predicted label was bounded by two opposite classifications, denoting an isolated classification. Therefore, in the post-processing, a sequence of 010 was corrected to 000 and 101 to 111.

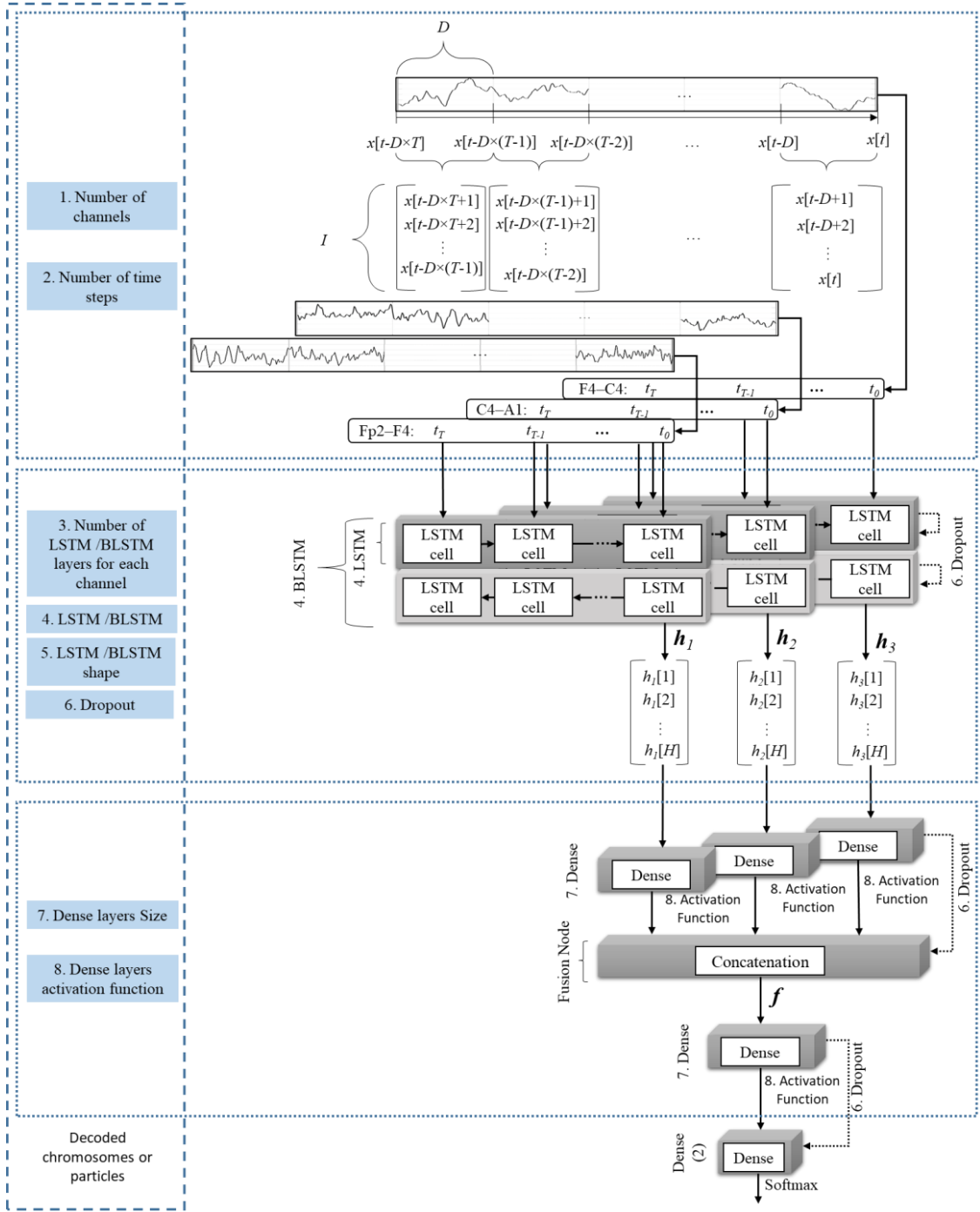

**Fig. 2.** Overview of the implemented model fusing the signal of three EEG channels, using a dense layer to transform the LSTM and concatenation layers outputs.

## 4. Experimental results

The algorithms were developed in Python 3 using TensorFlow's libraries to implement the classifier, running in NVIDIA's GeForce GTX 1080 Ti graphics processing unit. The first step was the search for the best structure of the classifier, performed by the optimization algorithms using TFCV. For the classifiers whose structure was found to be the best by the optimization algorithms, a second performance assessment was carried out by LOO (with a cold start of the classifier in each run).

**4.1 Optimization of the classifier**

The optimal parameters found by the optimization algorithms are presented in Table 2. Figures 3 and 4 present the AUC variation and the diversity of the chromosomes or particles through the evaluated generations or iterations, respectively. The simulation time was 1058067 seconds (12.25 days) and 859373 seconds (9.95 days) for the GA and PSO algorithms, respectively. A total of 300 different networks were simulated by GA while PSO simulated 255 different networks. It is important to notice that if a full grid search methodology was employed, the total number of examined networks would be 28672 which is computationally infeasible.

It was observed in Table 2 that both optimization algorithms identified a similar optimal structure, using the three EEG channels, a single BLSTM layer for each channel with the same shape, and employing the dense layers (one after each BLSTM layer and one after the concatenation layer). On the other hand, the chosen number of time steps was 25 for PSO, which was relatively higher when compared to GA which was 10, with a 10% lower dropout. The selected size and activation function for the dense layer was also different. The total number of trainable parameters was 934202 and 723602 for GA and PSO, respectively.

PSO was able to find the best solution at the second iteration, early stopping at iteration 16 (see Fig. 3). This could mean, however, that PSO converged prematurely, getting trapped into that local optimum. Nevertheless, it was significantly faster than GA, which reached the best solution at generation 15. PSO also maintained a higher diversity in the population (see Fig. 4). These results were expected as PSO is prone to converge faster while GA maintains the cycle of offspring creation which has the tendency to progressively decrease the diversity of the population.

**Table 2.**

Optimal configurations found by the optimization algorithms.

| Number | Parameters | Using GA | Using PSO |
|---|---|---|---|
| 1 | Number of channels to be fused | 3 (Fp2–F4, F4–C4, and C4–A1) | 3 (Fp2–F4, F4–C4, and C4–A1) |
| 2 | Number of time steps to be considered by the LSTM | 10 | 25 |
| 3 | Number of LSTM layers for each channel | 1 | 1 |
| 4 | Type of LSTM | BLSTM | BLSTM |
| 5 | Shape of the LSTM layers | 100 | 100 |
| 6 | Percentage of dropout for the recurrent and dense layers | 15% | 5% |
| 7 | Size of the dense layers | 300 | 200 |
| 8 | Activation function for the dense layers | Sigmoid | ReLu |

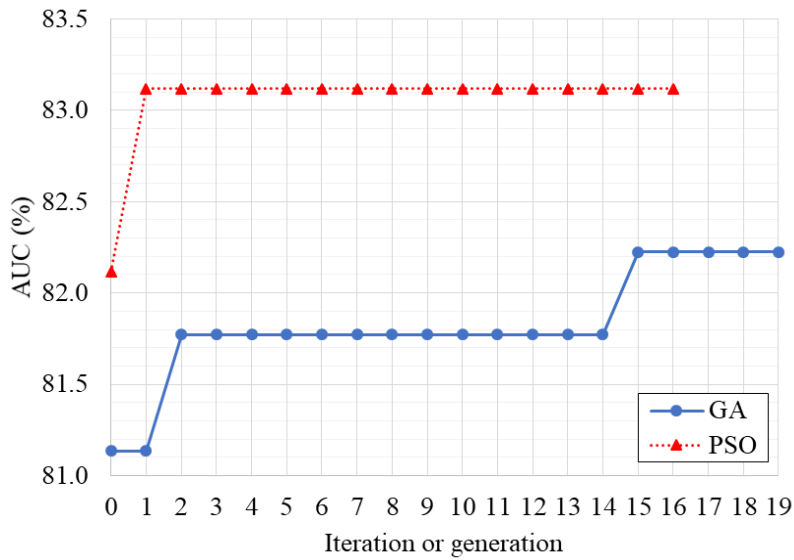

**Fig. 3.** Variation of the AUC of the best solution found by the optimization algorithms.

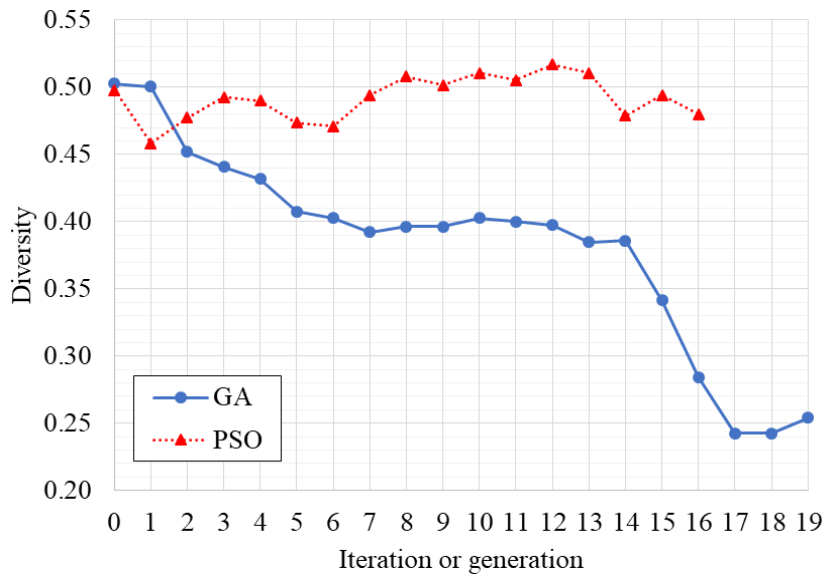

**Fig. 4.** Diversity of the chromosomes or particles over the optimization algorithms' iterations.

**4.2 Performance assessment**

The results obtained by the LOO method using the optimal configurations found by GA and PSO are presented in Tables 3, with the 16 subjects; with only the 8 subjects FND; with only the 8 subjects who have NFLE. Fig. 5 depicts the AUC for each of the subject (subjects 1 to 8 are FND while subjects 9 to 16 have NFLE).

By examining the results from Table 3, when the 16 subjects were used, it is possible to conclude that the configuration found by PSO reached an Acc and Spe which are approximately 3% and 4% better than the configuration found by GA, respectively. However, the results are less balanced when compared to the configuration found by GA which attained a Sen almost 5% higher. Nevertheless, the AUC of both configurations was approximately the same (82%), indicating that the performance of the two models is equivalent and that both optimization algorithms identified suitable configurations for this analysis. Another relevant aspect, highlighted in Fig. 5, is the variation of the performance according to the subjects, demonstrating that the models have an average absolute difference of 1%, and both are capable of working with subjects FND and subjects with NFLE, advocating the feasibility of the proposed model for clinical applications.

When comparing the LOO results (in Table 3) of the models using only the eight subjects FND or only the eight subjects which have NFLE against the LOO results with the 16 subjects, it is possible to observe that a superior performance for most performance metrics was reached when using LOO with the 16 subjects. These results were expected since the models were optimized to find the best solution when taking into consideration a population with both subjects FND and subjects with NFLE. Therefore, the proposed model has the key advantage of being capable of working with both a population FND and a population with sleep disorders (in this case with NFLE).

**Table 3.**

Performance attained by the LOO method for the best models identified by the optimization algorithms. Results are presented in the format "mean ± standard deviation (minimum value – maximum value)", and the studied population is referred to as FND and NFLE - Sleep Disorder Patients (SDP)

| Performance metric | Population (subjects) | Configuration found by GA | Configuration found by PSO |
|---|---|---|---|
| Acc (%) | 8 FND + 8 SDP | 76.52 ± 4.75 (68.08 – 85.30) | 79.43 ± 4.91 (69.25 – 87.29) |
|  | 8 FND | 76.53 ± 4.88 (70.67 – 87.01) | 77.24 ± 6.34 (69.16 – 86.16) |
|  | 8 SDP | 77.66 ± 4.55 (71.72 – 85.91) | 79.33 ± 4.74 (71.50 – 85.35) |
| Sen (%) | 8 FND + 8 SDP | 72.93 ± 9.77 (52.64 – 84.99) | 68.14 ± 11.26 (49.36 – 82.46) |
|  | 8 FND | 70.04 ± 9.67 (54.86 – 80.02) | 62.79 ± 12.79 (37.60 – 80.76) |
|  | 8 SDP | 70.67 ± 12.21 (51.73 – 85.12) | 65.14 ± 14.27 (43.46 – 85.51) |
| Spe (%) | 8 FND + 8 SDP | 77.07 ± 5.96 (66.69 – 88.12) | 81.21 ± 6.71 (68.79 – 93.35) |
|  | 8 FND | 77.28 ± 6.05 (69.65 – 89.22) | 79.02 ± 8.40 (67.90 – 91.95) |
|  | 8 SDP | 78.69 ± 6.60 (70.83 – 90.74) | 81.90 ± 7.10 (69.83 – 93.73) |
| AUC (%) | 8 FND + 8 SDP | 82.37 ± 4.75 (72.79 – 89.81) | 82.25 ± 4.53 (74.37 – 90.69) |
|  | 8 FND | 80.31 ± 4.67 (72.94 – 87.84) | 78.13 ± 3.89 (71.86 – 83.82) |
|  | 8 SDP | 82.26 ± 4.75 (74.16 – 89.52) | 81.69 ± 4.96 (74.54 – 91.10) |

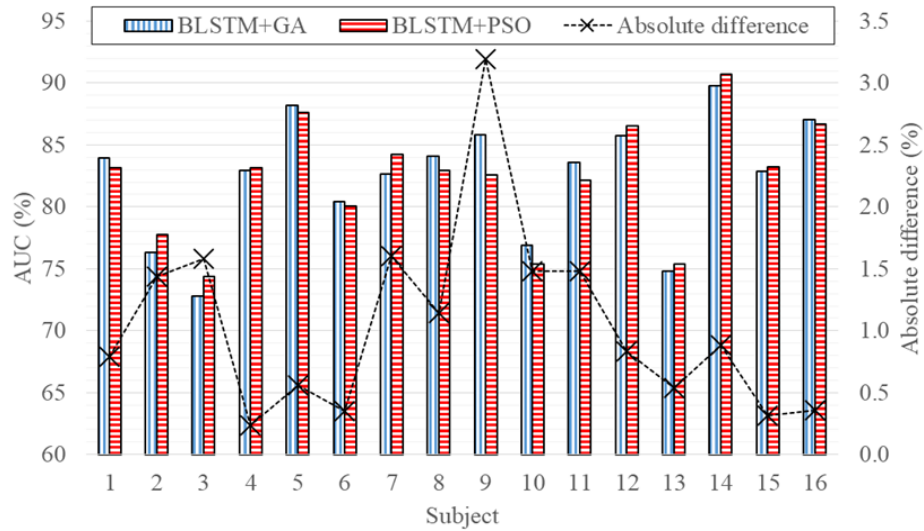

**Fig. 5.** AUC estimation using LOO for the models optimized by GA (BLSTM+GA) and PSO (BLSTM+PSO), depicting the absolute difference between the performance for each examined subject (model evaluating the 16 subjects).

### 4.3 Robustness evaluation

To evaluate the robustness of the proposed fusion method, two different tests were performed. The first examined the effect of losing the information from one or two channels, while the second evaluated the effect of introducing noise in the input signals. All results were attained by training the model with the three channels and without introducing noise, and then the models were tested with missing the channels or with the introduction of noise.

The results of the first test were attained by using LOO on the full population (16 subjects), covering all possible scenarios, and are presented in Fig. 6. For the scenario where no channels were lost indicating three (all) working channels, one channel was lost (indicated as two working channels in the figure) and two channels were lost (indicated as one working channel in the figure). The lost channel is replaced by one of the working channels or channel; as for two working channels it can be replaced by either one of them, where for one working channel all three channels' inputs are replaced by the remaining channel. By evaluating the results from Fig. 6 it is possible to conclude that losing one channel does not considerably change the AUC. Losing two channels (worst case scenario) decreased the AUC median less than 3% for both models, advocating the robustness of the models.

To evaluate the effect of having noise in the input signals, all EEG channels were contaminated with Additive White Gaussian Noise (AWGN) with varied Signal to Noise Ratio (SNR) from –20 to 20 dB (range considered as suitable for this type of analysis [73]). The results are presented in Fig. 7 where it is visible that the model whose structure was selected by GA is less affected by noise than the structure selected by PSO, conceivably due to the larger number of time steps used by the structure selected by PSO (15 time steps more than the structure selected GA), which means that more noise will affect the model. Nevertheless, both models maintained a good performance until the SNR was 0 dB, which is a value considerably lower than the usual SNR of EEG sensors [74]. Therefore, the proposed solution is also resistant to the introduction of noise in the input channels.

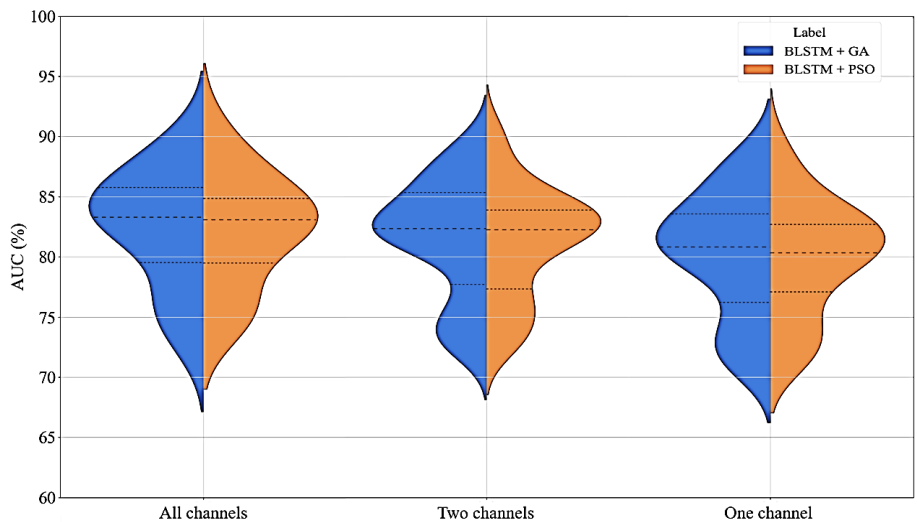

**Fig. 6.** Violin plots of the results attained by LOO when all three channels are available (indicated as "All channels"), when one channel failed (indicated as "Two channels"), and when two channels failed (indicated as "One channel"), for the models optimized

by GA (BLSTM+GA, in the left) and PSO (BLSTM+PSO, in the right), depicting the three quartiles (model evaluating the 16 subjects).

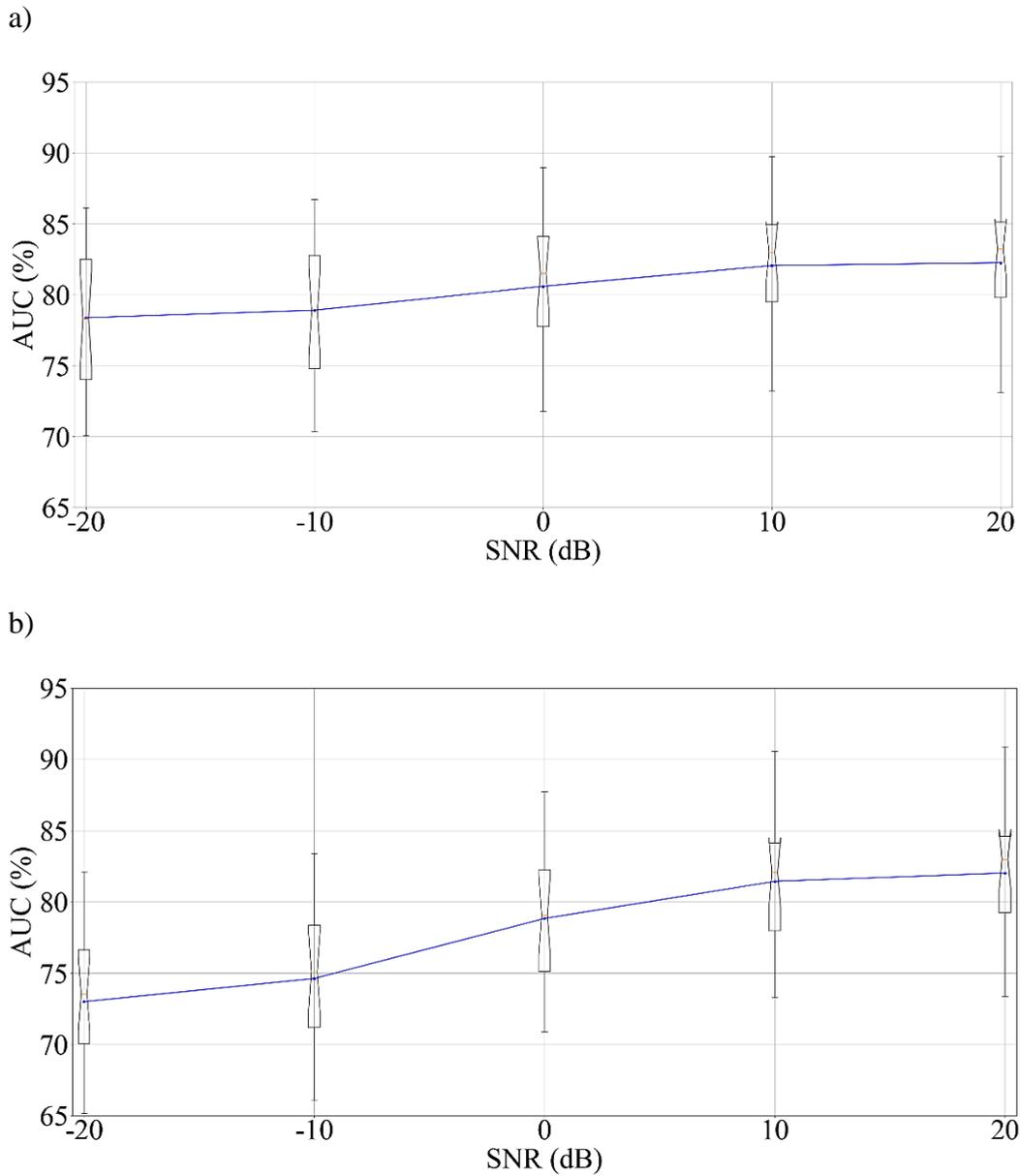

a)

b)

**Fig. 7.** Boxplots with the indication of the average values for the simulations where AWGN is introduced in the EEG signals, evaluating the 16 subjects using LOO, for the models optimized by a) GA, and b) PSO.

## 5. Discussion

A comparison between the results reported by the previous state of the art works and the results attained in this work is presented in Table 4. By examining the table, it is clear that the previous works which have only examined FNB subjects attained the best performance, highlighting the difficulties associated with the assessment of subjects with sleep disorders. Although the use of sleep disorder subjects made the classification process more challenging, the produced results can be better generalized for clinical applications. Another relevant factor is the average number of examined subjects by the state of the art works, which is 12, while 18 were examined in this work, emphasizing the viability of the achieved results. It is also important to highlight here the examination of multiple channels considering that, apart from Sharma et al. [43] who evaluated two EEG channels, all state of the art works examined only one EEG channel, which is contrary to the recommendation to score CAP utilizing multiple channels [38], given that an A phase can only be scored if it is visible in all EEG channels. The relevance of using multiple channels is even more emphasized in this work as both optimization algorithms selected as the best solution the use of three EEG channels.

Contrary to what was done in the developed models, most of the state of the art works have performed a manual removal of the wake or rapid eye movement periods [36] [39], which can boost the performance of the classifier, however it leads to a methodology which is not suitable to be implemented in a fully automatic scoring algorithm. Additionally, several state of the art works have further removed the epochs not related to the CAP phase events, further lessening the fully automatic applicability of the model [43].

For biomedical applications, it is important to have a balanced performance to provide a reliable clinical diagnosis. Taking into consideration the significant unbalance that characterizes CAP analysis (considerably more events related to "not-A" than "A"), it is not possible to focus the performance assessment only in the Acc since without reporting the Sen and Spe it is not possible to assess if the performance is balanced or not. For this reason, the AUC is preferable, although this metric was not reported by the majority of the state of the art works and, therefore, the mean metric was proposed in this analysis as an alternative to check how balanced the results are. By considering this metric, it is possible to conclude that the best state of the art works, which have included sleep disorder patients in the analysis, is the same as attained in this work (76%). However, Mendonça et al. [20] [41] have examined patients with sleep disordered breathing while subjects with NFLE were examined in this work. Sharma et al. [43] also evaluated subjects with NFLE but attained a lower Acc, highlighting how difficult it is to examine subjects with this disorder.

It is also important to notice that some state of the art works used a threshold based approach instead of a machine learning classifier [34] [35], which is likely to be difficult to generalize to a broader population. The works based on the manual creation of features to be fed to a classifier also have the disadvantage of requiring significant domain knowledge that hampers the researcher work [18]. Moreover, that methodology usually requires a feature selection procedure to determine the subset of features that are more relevant for the examined problem. On the other hand, the deep learning approach employed in this work automatically creates features, and can be further improved as more data is available, making the model more suitable for large scale examinations.

**Table 4.**

Comparative analysis between results reported by the state of the art works and the results attained in this work with subjects FND and Sleep Disorder Patients (SDP).

| Work | Population (subjects) | Examined channel | Acc (%) | Sen (%) | Spe (%) | Mean (%) |
|---|---|---|---|---|---|---|
| [46] | 15 FND | C4–A1 or C3–A2 | 70 | 51 | 81 | 67 |
| [36] | 8 FND | C4–A1 or C3–A2 | 72 | 52 | 76 | 67 |
| [34] | 6 FND | C4–A1 or C3–A2 | 81 | 76 | 81 | 79 |
| [32] | 12 FND* | - | 81 | 78 | 85 | 81 |
| [37] | 4 FND | C4–A1 or C3–A2 | 82 | 76 | 83 | 80 |
| [45] | 15 FND | C4–A1 or C3–A2 | 83 | 76 | 84 | 81 |
| [35] | 10 FND | F4–C4 | 84 | - | - | - |
| [38] | 4 FND | F4–C4 | 84 | 74 | 86 | 81 |
| [39] | 8 FND | C4–A1 or C3–A2 | 85 | 73 | 87 | 82 |
| [40] | 16 FND | C4–A1 or C3–A2 | 86 | 67 | 90 | 81 |
| [44] | 9 FND + 5 SDP | C4–A1 or C3–A2 | 67 | 55 | 69 | 64 |
| [43] | 27 SDP | C4–A1 and F4–C4 | 73 | - | - | - |
| [41] | 9 FND + 5 SDP | C4–A1 or C3–A2 | 75 | 78 | 74 | 76 |
| [20] | 15 FND + 4 SDP | C4–A1 or C3–A2 | 76 | 75 | 77 | 76 |
| Proposed BLSTM+GA | 8 FND +8 SDP | Fp2–F4, F4–C4, and C4–A1 | 77 | 73 | 77 | 76 |
| Proposed BLSTM+PSO | 8 FND +8 SDP | Fp2–F4, F4–C4, and C4–A1 | 79 | 68 | 81 | 76 |

* Evaluated one hour of data from each subject

## 6. Conclusion

A novel methodology to fuse time series signals at the feature level is proposed in this work, and it was evaluated in a challenging real-world scenario of CAP A phase classification. However, this methodology can be used in other contexts when it is intended to fuse information from multiple time series for classification or regression.

The proposed model automatically extracts features by identifying patterns in time from the input time series, using a deep learning classifier. However, one of the most challenging aspects of using deep learning models is the need to optimize the structure and hyperparameters. To address these problems, two optimization algorithms were examined as an efficient alternative to the traditional grid search approach to optimize. As a result, it was observed that the optimal structure for the classifier identified by the two optimization algorithms was similar and selected the input with three EEG signals, denoting the importance of using multiple channels to properly detect the CAP A phases.

It was observed that the obtained performance is in the upper range of the best state of the art works, although a significantly more challenging methodology and subjects data were employed in this work, examining a population composed of subjects FND and subjects with NFLE, using a fully automatic analysis instead of requiring to manually isolate the non-rapid eye movement sleep epochs as it is done in most of the state of the art works. It was also observed that the models are resilient to noise and channel failure, making them even more suitable for real-world clinical applications.

It is relevant to notice that the proposed architecture is flexible enough to be altered to include more layers (for example, a combination of convolution layer followed by an LSTM layer instead of only the LSTM layer) or to change the current layers (for example, change the LSTM to a gated recurrent unit).

Three main paths were identified as future work in this research. The first is to further validate the proposed methodology to include more channels in the analysis. The second one is to add different sensors to the fusion model. The last one is to implement a similar methodology to other research and industry applications.

**Acknowledgements**

Acknowledgment to the Portuguese Foundation for Science and Technology for their support through Projeto Estratégico LA 9 – UID/EEA/50009/2019.

Acknowledgment to ARDITI – Agência Regional para o Desenvolvimento da Investigação, Tecnologia e Inovação under the scope of the project M1420-09-5369-FSE-000002 – Post-Doctoral Fellowship, co-financed by the Madeira 14-20 Program - European Social Fund.

Acknowledgement to the Project MITIExcell (Project– UIDB/50009/2020), co-financed by Regional Development European Funds for the "Operational Programme Madeira 14–20" – Priority Axis 1 of the Autonomous Region of Madeira, number M1420-01-0145-FEDER-000002. Acknowledgement to the Project MTL – Marítimo Training Lab, number M1420-01-0247-FEDER-000033.

**CRediT authorship contribution statement**

Fábio Mendonça: Conceptualization, Methodology, Software, Implementation, Validation, Formal analysis, Investigation, Writing - Original Draft, Writing – review & editing.

Sheikh Shanawaz Mostafa: Conceptualization, Methodology, Software, Implementation, Validation, Formal analysis, Investigation, Writing - Original Draft, Writing – review & editing.

Diogo Freitas: Conceptualization, Methodology, Software, Implementation, Validation, Formal analysis, Investigation, Writing - Original Draft, Writing – review & editing.

Fernando Morgado-Dias: Conceptualization, Formal analysis, Investigation, Writing - Review & Editing, Supervision, Project administration, Funding acquisition.

Antonio G. Ravelo-García: Conceptualization, Formal analysis, Investigation, Writing - Review & Editing, Supervision, Project administration, Funding acquisition.

**Declaration of competing interest**

The authors declare that they have no interest that can influence the work reported in this article.

**References**

[1] B. Khaleghi, A. Khamis, F. Karray, S. Razavi, Multisensor data fusion: A review of the state-of-the-art, Information Fusion. 14 (2013) 28–44.

[2] S. Sun, H. Lin, J. Ma, X. Li, Multi-sensor distributed fusion estimation with applications in networked systems: A review paper, Information Fusion. 38 (2017) 122–134.

[3]  M. Fung, M. Chen, Y. Chen, Sensor Fusion: A Review of Methods and Applications, in: 2017 29th Chinese Control And Decision Conference (CCDC), Chongqing, China, 2017.

[4]  R. Gravina, P. Alinia, H. Ghasemzadeh, G. Fortino, Multi-sensor fusion in body sensor networks: State-of-the-art and research challenges, Information Fusion. 35 (2017) 68–80.

[5]  F. Mendonça, S. Mostafa, F. Morgado-Dias, A. Ravelo-García, Cyclic alternating pattern estimation based on a probabilistic model over an EEG signal, Biomedical Signal Processing and Control. 62 (2020) 102063.

[6]  M. Terzano, L. Parrino, A. Sherieri, R. Chervin, S. Chokroverty, C. Guilleminault, M. Hirshkowitz, M. Mahowald, H. Moldofsky, A. Rosa, R. Thomas, A. Walters, Atlas, rules, and recording techniques for the scoring of cyclic alternating pattern (CAP) in human sleep, Sleep Medicine. 2 (2001) 537–553.

[7]  M. Terzano, L. Parrino, Chapter 8 The cyclic alternating pattern (CAP) in human sleep, Handbook of Clinical Neurophysiology. 6 (2005) 79–93.

[8]  M. Terzano, D. Mancia, M. Salati, G. Costani, A. Decembrino, L. Parrino, The cyclic alternating pattern as a physiologic component of normal NREM sleep, Sleep. 8 (1985) 137–145.

[9]  P. Halász, M. Terzano, L. Parrino, R. Bódizs, The nature of arousal in sleep, Journal of Sleep Research. 13 (2004) 1–23.

[10] L. Parrino, G. Milioli, A. Melpignano, I. Trippi, The Cyclic Alternating Pattern and the Brain-Body-Coupling During Sleep, Epileptologie. 33 (2016) 150–160.

[11] L. Parrino, F. Ferrillo, A. Smerieri, M. Spaggiari, V. Palomba, M. Rossi, M. Terzano, Is insomnia a neurophysiological disorder? The role of sleep EEG microstructure, Brain Research Bulletin. 63 (2004) 377–383.

[12] L. Parrino, F. Paolis, G. Milioli, G. Gioi, A. Grassi, S. Riccardi, E. Colizzi, M. Terzano, Distinctive polysomnographic traits in nocturnal frontal lobe epilepsy, Epilepsia. 53 (2012) 1178–1184.

[13] M. Terzano, L. Parrino, M. Boselli, M. Spaggiari, G. Di Giovanni, Polysomnographic analysis of arousal responses in obstructive sleep apnea syndrome by means of the cyclic alternating pattern, Journal of Clinical Neurophysiology. 13 (1996) 145–155.

[14] L. Parrino, M. Boselli, P. Buccino, M. Spaggiari, G. Giovanni, M. Terzano, The cyclic alternating pattern plays a gate-control on periodic limb movements during non-rapid eye movement sleep, Journal of Clinical Neurophysiology. 13 (1996) 314–323.

[15] P. Halász, M. Terzano, L. Parrino, Décharges de pointes-ondes et microstructure du continuum veille-sommeil dans l'épilepsie généralisée idiopathique, Neurophysiologie Clinique. 32 (2002) 38–53.

[16] J. Rundo, R. Downey III, Chapter 25 - Polysomnography, in: Handbook of Clinical Neurology, Elsevier Science & Technology, Netherlands, 2019: pp. 381–392.

[17] A. Rosa, G. Alves, M. Brito, M. Lopes, S. Tufik, Visual and automatic cyclic alternating pattern (CAP) scoring: inter-rater reliability study, Arquivos de Neuro-Psiquiatria. 64 (2006) 578–581.

[18] S. Mostafa, F. Mendonça, A. Ravelo-Garcia, G. Juliá-Serdá, F. Morgado-Dias, Multi-Objective Hyperparameter Optimization of Convolutional Neural Network for Obstructive Sleep Apnea Detection, IEEE Access. 8 (2020) 129586–129599.

[19] T. Cover, The Best Two Independent Measurements Are Not the Two Best, IEEE Transactions on Systems, Man, and Cybernetics. SMC-4 (1974) 116–117.

[20] F. Mendonça, S. Mostafa, F. Morgado-Dias, A. Ravelo-García, A Portable Wireless Device for Cyclic Alternating Pattern Estimation from an EEG Monopolar Derivation, Entropy. 21 (2019) 1203.

[21] S. Albelwi, A. Mahmood, A framework for designing the architectures of deep convolutional neural networks, Entropy. 9 (2017) 242.

[22] F. Machado, C. Teixeira, C. Santos, C. Bento, F. Sales, A. Dourado, A-phases subtype detection using different classification methods, in: Florida, USA, 2016.

[23] E. Arce-Santana, A. Alba, M. Mendez, V. Arce-Guevara, A-phase classification using convolutional neural networks, Medical & Biological Engineering & Computing. 58 (2020) 1003–1014.

[24] M. Mendez, A. Alba, I. Chouvarda, G. Milioli, A. Grassi, M. Terzano, L. Parrino, On separability of A-phases during the cyclic alternating pattern, in: Chicago, USA, 2014.

[25] M. Mendez, I. Chouvarda, A. Alba, A. Bianchi, A. Grassi, E. Arce-Santana, G. Milioli, M. Terzano, L. Parrino, Analysis of A-phase transitions during the cyclic alternating pattern under normal sleep, Medical & Biological Engineering & Computing. 54 (2016) 133–148.

[26] I. Chouvarda, M. Mendez, A. Alba, A. Bianchi, A. Grassi, E. Arce-Santana, V. Rosso, M. Terzano, L. Parrino, Nonlinear Analysis of the Change Points between A and B phases during the Cyclic Alternating Pattern under Normal Sleep, in: 34th Annual International Conference of the IEEE EMBS, California, USA, 2012.

[27] S. Hartmann, M. Baumert, Automatic A-Phase Detection of Cyclic Alternating Patterns in Sleep Using Dynamic Temporal Information, IEEE Transactions on Neural Systems and Rehabilitation Engineering. 27 (2019) 1695–1703.

[28] C. Yeh, W. Shi, Generalized multiscale Lempel–Ziv complexity of cyclic alternating pattern during sleep, Nonlinear Dynamics. 93 (2018) 1899–1910.

[29] C. Navona, U. Barcaro, E. Bonanni, F. Martino, M. Maestri, L. Murri, An automatic method for the recognition and classification of the A-phases of the cyclic alternating pattern, Clinical Neurophysiology. 113 (2002) 1826–1831.

[30] F. Machado, F. Sales, C. Bento, A. Dourado, C. Teixeira, Automatic identification of Cyclic Alternating Pattern (CAP) sequences based on the Teager Energy Operator, in: Milan italy, 2015.

[31] C. Yeh, W. Shi, Identifying Phase-Amplitude Coupling in Cyclic Alternating Pattern using Masking Signals, Scientific Reports. 8 (2018) 2649.

[32] R. Largo, C. Munteanu, A. Rosa, CAP Event Detection by Wavelets and GA Tuning, in: Faro, Portugal, 2005.

[33] R. Largo, C. Munteanu, A. Rosa, Wavelet based CAP detector with GA tuning, in: Lisbon, Portugal, 2005.

[34] H. Niknazar, S. Seifpour, M. Mikaili, A. Nasrabadi, A. Banaraki, A Novel Method to Detect the A Phases of Cyclic Alternating Pattern (CAP) Using Similarity Index, in: Tehran, Iran, 2015.

[35] U. Barcaro, E. Bonanni, M. Maestri, L. Murri, L. Parrino, M. Terzano, A general automatic method for the analysis of NREM sleep microstructure, Sleep Medicine. 5 (2004) 567–576.

[36] S. Mariani, E. Manfredini, V. Rosso, M. Mendez, A. Bianchi, M. Matteucci, M. Terzano, S. Cerutti, L. Parrino, Characterization of A phases during the Cyclic Alternating Pattern of sleep, Clinical Neurophysiology. 122 (2011) 2016–2024.

[37] S. Mariani, A. Bianchi, E. Manfredini, V. Rosso, M. Mendez, L. Parrino, M. Matteucci, A. Grassi, S. Cerutti, M. Terzano, Automatic detection of A phases of the Cyclic Alternating Pattern during sleep, in: Buenos Aires, Argentina, 2010.

[38] S. Mariani, A. Grassi, M. Mendez, L. Parrino, M. Terzano, A. Bianchi, Automatic detection of CAP on central and fronto-central EEG leads via Support Vector Machines, in: 33rd Annual International Conference of the IEEE Engineering in Medicine and Biology Society, Massachusetts, USA, 2011.

[39] S. Mariani, E. Manfredini, V. Rosso, A. Grassi, M. Mendez, A. Alba, M. Matteucci, L. Parrino, M. Terzano, S. Cerutti, A. Bianchi, Efficient automatic classifiers for the detection of A phases of the cyclic alternating pattern in sleep, Medical & Biological Engineering & Computing. 50 (2012) 359–372.

[40] S. Mariani, A. Grassi, M. Mendez, G. Milioli, L. Parrino, M. Terzano, A. Bianchi, EEG segmentation for improving automatic CAP detection, Clinical Neurophysiology. 124 (2013) 1815–1823.

[41] F. Mendonça, A. Fred, S. Mostafa, F. Morgado-Dias, A. Ravelo-García, Automatic Detection of a Phases for CAP Classification, in: 7th International Conference on Pattern Recognition Applications and Methods (ICPRAM), Funchal, Portugal, 2018.

[42] F. Mendonça, A. Fred, S. Mostafa, F. Morgado-Dias, A. Ravelo-García, Automatic detection of cyclic alternating pattern, Neural Computing and Applications. (2018) 1–11. https://doi.org/10.1007/s00521-018-3474-5.

[43] M. Sharma, V. Patel, J. Tiwari, U. Acharya, Automated Characterization of Cyclic Alternating Pattern Using Wavelet-Based Features and Ensemble Learning Techniques with EEG Signals, Diagnostics. 11 (2021) 1380.

[44] S. Mostafa, F. Mendonça, A. Ravelo-García, F. Morgado-Dias, Combination of Deep and Shallow Networks for Cyclic Alternating Patterns Detection, in: 2018 13th APCA International Conference on Automatic Control and Soft Computing (CONTROLO), Ponta Delgada, Portugal, 2018.

[45] S. Hartmann, M. Baumert, Improved A-phase Detection of Cyclic Alternating Pattern Using Deep Learning, in: 2019 41st Annual International Conference of the IEEE Engineering in Medicine and Biology Society (EMBC), Berlin, Germany, 2019.

[46] F. Mendonça, S. Mostafa, F. Morgado-Dias, A. Ravelo-Garcia, Cyclic Alternating Pattern Estimation from One EEG Monopolar Derivation Using a Long Short-Term Memory, in: 2019 International Conference in Engineering Applications (ICEA), São Miguel, Portugal, 2019.

[47] M. Ravan, J. Begnaud, Investigating the Effect of Short Term Responsive VNS Therapy on Sleep Quality Using Automatic Sleep Staging, IEEE Transactions on Biomedical Engineering. 66 (2019) 3301–3309.

[48] A. Goldberger, L. Amaral, L. Glass, M. Hausdorff, P. Ivanov, R. Mark, J. Mietus, G. Moody, C. Peng, H. Stanley, PhysioBank, PhysioToolkit, and PhysioNet: Components of a new research, Circulation. 101 (2000) 215–220.

[49] F. Gers, J. Schmidhuber, F. Cummins, Learning to forget: Continual prediction with LSTM, Neural Computation. 12 (2000) 2451–2471.

[50] I. Goodfellow, Y. Bengio, A. Courville, Deep Learning, The MIT Press, Massachusetts, 2016.

[51] S. Panda, N. Padhy, Comparison of particle swarm optimization and genetic algorithm for FACTS-based controller design, Applied Soft Computing. 8 (2008) 1418–1427.

[52] C. Jennison, N. Sheehan, Theoretical and Empirical Properties of the Genetic Algorithm as a Numerical Optimizer, Journal of Computational and Graphical Statistics. 4 (1995) 296–318.

[53] Y. Fang, J. Li, A Review of Tournament Selection in Genetic Programming, in: Advances in Computation and Intelligence - 5th International Symposium, ISICA 2010, Wuhan, China, 2010.

[54] D. Hakimi, D. Oyewola, Y. Yahaya, G. Bolarin, Comparative Analysis of Genetic Crossover Operators in Knapsack Problem, Journal of Applied Sciences and Environmental Management. 20 (2016) 593–596.

[55] M. Črepinšek, S. Liu, M. Mernik, Exploration and Exploitation in Evolutionary Algorithms: A Survey, ACM Computing Surveys. 45 (2013) 1–33.

[56] R. Eberhart, J. Kennedy, A New Optimizer Using Particle Swarm Theory, in: 6th International Symposium on Micro Machine and Human Science, 6th International Symposium on Micro Machine and Human Science, Nagoya, Japan, 1995: pp. 39–43.

[57] J. Kennedy, R. Eberhart, Particle Swarm Optimization, in: International Conference on Neural Networks, International Conference on Neural Networks, Perth, Australia, 1995: pp. 1942–1948.

[58] J. Kennedy, R.C. Eberhart, A Discrete Binary Version of the Particle Swarm Algorithm, in: IEEE International Conference on Systems, Man and Cybernetics, IEEE International Conference on Systems, Man and Cybernetics, Florida, USA, 1997: pp. 4104–4108.

[59] Y. Shi, R.C. Eberhart, A Modified Particle Swarm Optimizer, in: IEEE World Congress on Computational Intelligence, IEEE World Congress on Computational Intelligence, Alaska, USA, 1998: pp. 69–73.

[60] J. Kennedy, R. Mendes, Population Structure and Particle Swarm Performance, in: IEEE Congress on Evolutionary Computation, IEEE Congress on Evolutionary Computation, Hawaii, USA, 2002: pp. 1671–1676.

[61] D. Bratton, J. Kennedy, Defining a Standard for Particle Swarm Optimization, in: IEEE Swarm Intelligence Symposium, IEEE Swarm Intelligence Symposium, Hawaii, USA, 2007: pp. 120–127.

[62] D. Sackett, R. Haynes, G. Guyatt, P. Tugwell, Clinical Epidemiology: A Basic Science for Clinical Medicine, 2nd ed., Lippincott Williams and Wilkins, Pennsylvania, USA, 1991.

[63] T. Fawcett, An introduction to ROC analysis, Pattern Recognition Letters. 27 (2006) 861–874.

[64] N. Shamir, D. Saad, E. Marom, Preserving the Diversity of a Genetically E volving Population of Nets U sing the Functional Behavior of Neurons, Complex Systems. 7 (1993) 327–346.

[65] D. Kingma, J. Ba, Adam: A Method for Stochastic Optimization, in: California, USA, 2015.

[66] R. Kohavi, A study of cross-validation and bootstrap for accuracy estimation and model selection, in: Quebec, Canada, 1995.

[67] I. Digital Signal Processing Committee, Programs for Digital Signal Processing, IEEE press, New York, USA, 1979.

[68] K. Muralidharan, A Note on Transformation, Standardization and Normalization, The IUP Journal of Operations Management. 9 (2010) 116–122.

[69] J. Urigüen, B. Zapirain, EEG artifact removal – State-of-the-art and guidelines, Journal of Neural Engineering. 12 (2015) 031001.

[70] K.R. Harrison, A.P. Engelbrecht, B.M. Ombuki-Berman, Optimal parameter regions and the time-dependence of control parameter values for the particle swarm optimization algorithm, Swarm and Evolutionary Computation. 41 (2018) 20–35.

[71] R.C. Eberhart, Y. Shi, Comparing Inertia Weights and Constriction Factors in Particle Swarm Optimization, in: IEEE Congress on Evolutionary Computation, IEEE Congress on Evolutionary Computation, California, USA, 2000: pp. 84–88.

[72] J. Xin, G. Chen, Y. Hai, A Particle Swarm Optimizer with Multi-Stage Linearly-Decreasing Inertia Weight, in: 2nd International Joint Conference on Computational Sciences and Optimization, 2nd International Joint Conference on Computational Sciences and Optimization, Sanya, China, 2009: pp. 505–508.

[73] M. Kwon, S. Han, K. Kim, S. Jun, Super-Resolution for Improving EEG Spatial Resolution using Deep Convolutional Neural Network—Feasibility Study, Sensors. 19 (2019) 5317.

[74] M. O'Sullivan, A. Temko, A. Bocchino, C. O'Mahony, G. Boylan, E. Popovici, Analysis of a Low-Cost EEG Monitoring System and Dry Electrodes toward Clinical Use in the Neonatal ICU, Sensors. 19 (2019) 2637.